\documentclass[a4paper,11pt]{article}
\usepackage[dvips]{graphicx}
\usepackage{amssymb}
\usepackage{amsmath}

\usepackage{cite}

\begin{document}

\title{An Electric Analogue to Gravity Induced Vacuum Dominance}
\author{V. K. Oikonomou\thanks{
voiko@physics.auth.gr}\\
Department of Theoretical Physics, Aristotle University of Thessaloniki,\\
Thessaloniki 541 24 Greece} \maketitle

\begin{abstract}
It is shown that when a strong electric field is turned on as a
background in a charged scalar field, the vacuum state of the
scalar field becomes exponentially amplified. This is an analogue
to gravity induced vacuum amplification, where the vacuum state
becomes exponentially amplified for some geometries.
\end{abstract}

\bigskip

\bigskip

In this letter we intend to show the analogy
between quantum field theory in extreme gravitational environments
and quantum field theory in strong electric fields. The
similarities that the aforementioned theories have, were pointed
out long ago \cite{fulling}, but not in a quantitative way. We will demonstrate that the vacuum energy
of a charged scalar field has an exponential time dependence, when
strong electric fields (in flat spacetime) are applied. The
exponential amplification also occurs when one considers scalar fields in gravitational fields \cite{lima,limasos,lima1}.

\noindent The usual particle interpretation collapses due to the
exponential factor, however it is exactly this that makes this
phenomenon so interesting. This phenomenon is closely related to
the definition of vacuum state, and in general to the re-definition
of the Fock space of the quantum system. It is probable that
another definition of the vacuum state could alter this
exponential amplification, but we shall not consider this problem
here \cite{ss}.

\noindent We begin our presentation by describing the
gravitational vacuum dominance pointed out by Lima et al.
\cite{lima,limasos,lima1}. They considered a free scalar field, with
a coupling to scalar curvature $\xi$. The spacetime $(M,g^{\mu\nu})$, $\mu,\nu=0,..4$, is considered to be globally hyperbolic, foliated with Cauchy surfaces $\Sigma_t$. These surfaces are covered with the coordinates $\vec{x}=(x^1,x^2,x^3)$, satisfying $n^\mu \nabla_{\mu}x^i=0$, with $\mu =0,..3$ and $i=1,2,3$. The general metric of the spacetime can be written in the form:
\begin{equation}\label{met1}
\mathrm{d}s^2=N^2(-\mathrm{d}t^2+h_{ij}(t,x^i){\,}\mathrm{d}x^i\mathrm{d}x^j )
\end{equation}
with $N=N(t,x^i)>0$ and $g_{ij}(t,x^i)=N^2h_{ij}(t,x^i)$ is the three dimensional spatial metric induced on each Cauchy surface $\Sigma_t$. The dynamics of the scalar field in the spacetime is governed by the action:
\begin{equation}\label{saction}
\mathcal{S}=-\frac{1}{2}\int_M \mathrm{d}x^4\sqrt{-g}((\nabla _{\mu}\Phi)^*\nabla ^{\mu}\Phi+m^2|\Phi|^2+\xi R|\Phi|^2)
\end{equation}
with $R$, the Ricci scalar. The scalar field satisfies
the Klein-Gordon equation,
\begin{equation}\label{confscalar}
\left (-\Box+m^2+\xi R  \right )\Phi=0
\end{equation}
The Klein-Gordon inner product between any two solutions $u,v$ of the above equation is:
\begin{equation}\label{innerpr}
(u,v)_{KG}=i\int_{\Sigma_t}\mathrm{d}\Sigma n^{\mu}[u^*\nabla_{\mu} v-v\nabla_{\mu} u^*],
\end{equation}
with $\mathrm{d}\Sigma$ a proper volume element on $\Sigma_t$, and the inner product being independent from the choice of $\Sigma_t$. The scalar field satisfies the equal-time canonical commutation relations that stem from
the quantization of the field in curved spacetime,
\begin{equation}\label{ccreghddff}
[\hat{\Phi}(t,\vec{x}),\hat{\Pi}(t,\vec{x}')]_{\Sigma_t}=i\delta^3 (\vec{x},\vec{x}')
\end{equation}
with $\vec{x}=(x^1,x^2,x^3)$. A representation of these commutation relations, can be materialized by using the positive and negative norm solutions of the equation (\ref{confscalar}), namely $u_a^{(+)}$ and $u_a^{(-)}=(u_a^{(+)})^*$, which together form a complete set of normal modes that satisfy:
\begin{equation}\label{ghcghfgh}
(u_a^{(+)},u_b^{(+)})=-(u_a^{(-)},u_b^{(-)})=\delta (a,b),{\,}{\,}{\,}(u_a^{(+)},u_b^{(-)})=0
\end{equation}
In the above, $(a,b)$ represent all the quantum numbers of the scalar field, and $\delta (a,b)$ is the delta function associated with the quantum numbers $(a,b)$. The scalar field operator can be written in terms of the positive and negative norm functions, as follows:
\begin{equation}\label{hfgjkilf}
\Phi=\int \mathrm{d}\mu (\sigma)[a_{\sigma}u_{\sigma}^{(+)}+a_{\sigma}^{\dag}u_{\sigma}^{(-)}]
\end{equation}
where $\mu$ is a measure defined on the set of quantum numbers, denoted as ``$\sigma$''. As a consequence of 
(\ref{ccreghddff}), the annihilation and creation operators $a_{\sigma},a_{\sigma}^{\dag}$ satisfy the commutation relations:
\begin{equation}\label{commmmtrr}
[a_{\sigma},a_{\sigma '}^{\dag}]=\delta (\sigma,\sigma ')
\end{equation}
The vacuum state $\rvert 0\rangle$, is defined by requiring,
\begin{equation}\label{bvnnbbbvnvb}
a_{\sigma} \rvert 0\rangle =0
\end{equation}
The Fock space is based on this vacuum definition, the existence of which depends strongly on the commutation relations (\ref{ccreghddff}).

\noindent We assume that the metric in the asymptotic past (in) and in the asymptotic future (out) is conformally static, and particularly flat in the asymptotic past,
\begin{equation}\label{coon}
\mathrm{d}s^2\sim \Bigg{\{}\begin{array}{c}
   f_{in}^2(t,\vec{x})\left(-\mathrm{d}t^2+\mathrm{d}x^2\right ),{\,}{\,}\mathrm{Past} \\
   \\
  f_{out}^2(t,\vec{x})\left(-\mathrm{d}t^2+h_{ij}{\,}\mathrm{d}x^i\mathrm{d}x^j\right ),{\,}{\,}\mathrm{Future}\\
\end{array}
\end{equation}
with $f_{out}(t,\vec{x})$ and $f_{in}(t,\vec{x})$, smooth functions of $t$ and $\vec{x}$, and
$h_{ij}=h_{ij}(\vec{x})$, with $\vec{x}=(x^1,x^2,x^3)$. In these two asymptotic regions, and by rescaling the field as $\tilde{\Phi}=\Phi/f_J$ ($J=$out,in), the equation of motion becomes:
\begin{equation}\label{sceqmot}
-\frac{\partial^2}{\partial t^2}\tilde{\Phi}=-\Delta_J\tilde{\Phi}+V_J\tilde{\Phi}
\end{equation}
In the above, $\Delta_J$ is the Laplace operator in the corresponding asymptotic regions, and the term $V_J$ is equal to:
\begin{equation}\label{pot}
V_J=(1-6\xi )\frac{\Delta_Jf_J-\frac{\mathrm{d}^2{f_J}}{\mathrm{d}t^2}}{f_J}+f_J^2m^2+\xi K_J
\end{equation}
with $K_{in}=0$ and $K_{out}=K(\vec{x})$ the scalar curvature associated with the spatial metric $h_{ij}$. Furthermore, we assume that $V_{out}=V(\vec{x})$, which means it is independent of time. With this assumption, there are two different sets of positive norm modes that are solutions of the Klein-Gordon equation of motion (\ref{confscalar}),
\begin{equation}\label{thecoon}
\begin{array}{c}
   u_{\vec{k}}^{+}\sim(16\pi^3\omega_{\vec{k}})^{-1/2}f_{in}^{-1}(t,\vec{x})e^{-i(\omega_{\vec{k}}t-{\vec{k}}x)} {\,}{\,}(\mathrm{Past})\\
   \\
  v_a^{(+)}\sim(2\omega_a)^{-1/2}f_{out}^{-1}(t,\vec{x})e^{-i\omega_at}F_a(\vec{x}){\,}{\,}(\mathrm{Future})\\
\end{array}
\end{equation}
with $\vec{k}$ $\in$ $R^3$, $\omega_a>0$. The indices ``$\vec{k}$'' and ``$a$'', stand for the wavenumber in the conformally flat asymptotic past spacetime and the quantum numbers of the scalar boson in the $h_{ij}$ spacetime, respectively. The function $F_a(\vec{x})$, is a solution of,
\begin{equation}\label{mysyre}
\left [-\Delta_{out}+V_{out}(\vec{x}) \right ]F_a(\vec{x})=\omega_a^2F_a(\vec{x})
\end{equation}
and satisfies the normalization:
\begin{equation}\label{norma}
\int_{\Sigma_{out}}\mathrm{d}x^3\sqrt{h}F^*_a(\vec{x})F_b(\vec{x})=\delta (a,b)
\end{equation}
on a surface $\Sigma_{out}$, in the asymptotic future. As it is obvious, the function $F_a(\vec{x})$ can be complex.
Whenever the above equation has imaginary eigenvalue solutions, $\omega_a^2=-\lambda_a^2$, the modes $v_a^{(+)}$ and $v_a^{(-)}=(v_a^{(+)})^*$ fail to form a complete set of normal modes. In this case, in order to have a complete basis in the space of solutions, we have to add additional positive and negative norm modes $w_a^{(+)}$ and $w_a^{(-)}=(w_a^{(+)})^*$, of the form, 
\begin{equation}\label{comcoon}
w_a^{(+)}\sim\frac{\left(e^{\lambda_at-i\pi/2}+e^{-\lambda_at+i\pi/2}\right
)F_a(\vec{x})}{\sqrt{2\lambda_a}f_{out}(t,\vec{x})} {\,}{\,}{\,}{\,}(\mathrm{future})
\end{equation}
corresponding to the imaginary eigenvalues $\omega_a^2=-\lambda_a^2$. The difference of the modes $w_a^{(+)}$ and $v_a^{(+)}$, is that the latter are eigenstates of the Killing vector field $\partial_t^2$, associated with the spacetime (\ref{met1}). Consequently and importantly the vacuum and the other Fock space states do not have in general any particle content interpretation \cite{lima,limasos}. This is a direct consequence of the existence of imaginary eigenvalues $\omega_a$. As we shall see in the electric field case, analogous arguments make the particle interpretation impossible. Turning back to the normal modes, the space of states corresponding to the asymptotic future is constituted by the four functions $v_a^{(+)},v_a^{(-)},w_a^{(+)},w_a^{(-)}$ and consequently, the field can be expanded in terms of these normal modes as follows:
\begin{equation}\label{hfgjkilf}
\Phi=\int \mathrm{d}\mu (\sigma)[b_{\sigma}v_{\sigma}^{(+)}+b_{\sigma}^{\dag}v_{\sigma}^{(-)}+c_{\sigma}w_{\sigma}^{(+)}+c_{\sigma}^{\dag}w_{\sigma}^{(-)}]
\end{equation}
The sets of operators $b_{\sigma}$, $b_{\sigma}^{\dag}$, and $c_{\sigma}$, $c_{\sigma}^{\dag}$ corresponding to the functions $v_{\sigma}$ and $w_{\sigma}$, are assumed to satisfy the following commutation relations:
\begin{equation}\label{commutationgravity}
\left [b_{\sigma},b_{\sigma '}^{\dag}\right
]=\delta_{\sigma \sigma'},{\,}{\,}{\,}{\,}{\,}{\,}{\,}\left
[c_{\sigma},c_{\sigma'}^{\dag}\right ]=\delta_{\sigma \sigma'}
\end{equation}
The commutation relations $\left [b_{\sigma},b_{\sigma '}^{\dag}\right
]=\delta_{\sigma \sigma'}$, naturally follow from the quantization conditions of the scalar field. However, the commutation relations $[c_{\sigma},c_{\sigma'}^{\dag} ]=\delta_{\sigma \sigma'}$, are not a result of the field quantization procedure. Actually the presence of the imaginary eigenvalues poses difficulties that can be solved in two ways \cite{ss,wolfram}:

\begin{itemize}
 \item Either define indefinite metric in terms of the commutation relations of the operators $c_{\sigma}$.
\item Or abandon the usual quantization relations.

\end{itemize}

As a consequence of the second way, the usual particle interpretation that the theory provides, collapses. In addition, the Hamiltonian of the system is unbounded from below, because it contains terms that are of the form of inverted harmonic oscillators, (always in terms of the operators $c_{\sigma}$, $c_{\sigma'}^{\dag}$) as can be seen in \cite{limasos}. The authors of \cite{lima,limasos,lima1} used the second approach and this is what we shall adopt in the electric field case too. Remarkably, the same features that appear in the gravitational scalar boson system, appear in the electric field scalar boson case too. 

\noindent The asymptotic past positive norm states $u_{\vec{k}}$, can be written in terms of the asymptotic future normal modes, $v_a^{(+)},v_a^{(-)},w_a^{(+)},w_a^{(-)}$. As a consequence of the existence of imaginary eigenvalues, some modes of the asymptotic past will undergo a phase of exponential growth, a fact that has its imprint on the vacuum energy. Indeed the vacuum energy of the field reads (calculated for the ``in" states),
\begin{equation}\label{gravvac}
 _{in}\langle 0\rvert \mathcal{T}^{00}\rvert 0\rangle_{in}\sim
\frac{\kappa e^{2\lambda_at}}{2\lambda_a}\left(\frac{F_a(\vec{x})}{f_{out}(t,\vec{x})}
\right)^2G(\xi,F_a(\vec{x}),\lambda_a,f_{out}(t,\vec{x}))[1+\mathcal{O}(e^{-\epsilon t})]
\end{equation}
In the above equation (\ref{gravvac}), the parameter $\epsilon$ is some positive constant and $\kappa$ is a dimensionless constant that depends on the spacetime structure and on the initial state. Actually it is determined by the outcome of the projection of each $u_{\vec{k}}^{(+)}$ on the imaginary eigenvalue mode $w_a^{(+)}$. Moreover, the function $G(\xi,F_a(\vec{x}),\lambda_a,f_{out}(t,\vec{x}))$, is actually equal to:
\begin{align}\label{fgfunxction}
&G(\xi,F_a(\vec{x}),\lambda_a,f_{out}(t,\vec{x}))=\frac{(1-4\xi)}{2}\Big{(}\lambda_a^2+\frac{(DF_a(\vec{x}))^2}{F_a(\vec{x})^2}+m^2f^2_{out}(t,\vec{x})+\xi K_{out}\Big{)}
\\ \notag & +(1-6\xi )\Big{(}\frac{ \frac{\mathrm{d}^2f_{out}(t,\vec{x})}{\mathrm{d}t^2}}{f_{out}(t,\vec{x})}+\frac{\Big{(}\frac{\mathrm{d}f_{out}(t,\vec{x})}{\mathrm{d}t}\Big{)}^2}{2f^2_{out}(t,\vec{x})}-\frac{\lambda_a
 \frac{\mathrm{d}f_{out}(t,\vec{x})}{\mathrm{d}t}}{f_{out}(t,\vec{x})}\Big{)}
+\\ \notag & +\frac{(Df_{out}(t,\vec{x}))^2}{2f_{out}(t,\vec{x}))^2}-\frac{Df_{out}(t,\vec{x})DF_a(\vec{x})}{f_{out}(t,\vec{x})F_a(\vec{x})}
\end{align}
where $D$ is defined so that $\Delta_{out}=D^2$. Thus the classical gravitational background exponentially amplifies the quantum field's vacuum energy. A reasonable question to ask is when equation (\ref{mysyre}) has imaginary eigenvalues as solutions. As pointed out in references \cite{lima,limasos}, the imaginary eigenvalues case occurs when $V_{out}$ gets sufficiently negative. This requirement as we shall see holds also true for the electric field case. It worths mentioning the physical requirements that must be satisfied, in order we have this phenomenon of exponential growth in the vacuum energy. As pointed out in \cite{lima,limasos,lima1}, for massless scalar field with the coupling $\xi$ being of order unity, the $V_{out}$ must be of order $R$ (the scalar curvature). Hence the background geometries associated with mass distributions having density variation $\delta\rho_c$ over regions with typical length $L$, must satisfy $8\pi G \delta\rho_c L^2 \sim 1$. The parameter ``$G$'' is Newton's constant.

%%%%%%%%%%%%%%%%%%%%%%%%%%%%%%%%%%%%%%%%%%%%%%%%%%_electric_part

\noindent Let us now turn our focus in the electric field vacuum amplification case. The quantum vacuum amplification phenomenon, as we already mentioned, appears in the case we have a scalar boson field in strong classical static external electric field. Consider a massive charged scalar
field with a static background field $\mathcal{A}^{\mu}(\vec{x})$. The Lagrangian is equal
to:
\begin{equation}\label{laplacian}
\mathcal{L}_M=\left[\Big{(}\partial^{\mu}-ie\mathcal{A}^{\mu}(\vec{x})\Big{)}\varphi\right]^*\left[\Big{(}\partial_{\mu}-ieA_{\mu}(\vec{x})\Big{)}\varphi\right]-m^2\varphi^2
\end{equation}
We take into account only the electric field $\mathcal{A}_{0}(\vec{x})$. The Klein-Gordon equation with a time-independent external electrostatic potential after separation of the time variable, can be written,
\begin{equation}\label{klein1}
\left (-\nabla^2+m^2\right )\phi_j(\vec{x})=\left (
\omega_j+e\mathcal{A}_{0}(\vec{x})\right )^2\phi_j(\vec{x})
\end{equation}
where we assumed that the solution takes the form $\varphi(\vec{x},t)=e^{-i\omega t}\phi_j(\vec{x})$. Complex eigenvalues of the above equation occur in square well electrostatic potentials \cite{fulling,ssw}. Imagine that the electric field $\mathcal{A}_{0}(\vec{x})$, takes the following form in space,
\begin{equation}
\mathcal{A}_{0}(\vec{x})= \Bigg{\{}\begin{array}{c}
   0, {\,}{\,} {\,}{\,}-L_{-}<\vec{x}<0 \\
   \\
  -V_0,{\,}{\,}{\,}{\,}{\,}{\,}0<\vec{x}<L_+\\
\end{array}
\end{equation}
with $L_+,L_{-}$, finite lengths. As was shown in \cite{fulling,ssw}, for sufficiently strong electric fields (deep potentials, which means large values of $V_0$), imaginary eigenfrequencies occur in the system. Actually for potentials for which the depth $V_0$ is larger than the threshold for pair creation, that is $|eV_0|\geq 2 m$ (with ``$m$'' the mass of the particle), the existence of imaginary eigenvalues in unavoidable. For imaginary eigenvalues of the form, $\omega_j^2=-\lambda_j^2$ the field expansion contains terms corresponding to real and imaginary eigenvalues \cite{ss,wolfram},
\begin{align}\label{expansion}
&\Phi(\vec{x},t)=\sum_na_ne^{-\lambda_nt}\phi_n(\vec{x})+b_ne^{\lambda_nt}\phi_n^*(\vec{x}) \\ \notag &
+\sum_{E_i\geq 0}c_ie^{-E_it}\Phi_i(\vec{x})+d_i^{\dag}e^{E_it}\Phi_i^*(\vec{x}) \\ \notag &
+\int c(k)e^{-i\omega_k t}\phi_k(\vec{x})+d^{\dag}(k)e^{E_it}\phi_i^*(\vec{x}) 
\end{align}
For the continuum energy and for the positive eigenfrequency discrete states, a consistent second quantization with canonical commutation relations:
\begin{equation}\label{cancom}
\left [\Phi(\vec{x},t),\pi (\vec{y},t) \right ]=i\delta^3 (\vec{x}-\vec{y})
\end{equation}
implies the following commutation relations for the operators $c_i,d_i$,
\begin{equation}\label{commutation12}
\left [c_i,c^{\dag}_j\right
]=\delta_{ij}{\,}{\,}{\,}{\,}{\,}{\,}{\,}\left
[d_i,d^{\dag}_j\right ]=\delta_{ij}
\end{equation}
Here is the crucial step. As the authors of \cite{lima,limasos,lima1} did in these papers, we also assume that the operators $a_n$ and $b_n$ satisfy the usual
commutation relations,
\begin{equation}\label{commutation}
\left [a_n,a^{\dag}_m\right
]=\delta_{nm}{\,}{\,}{\,}{\,}{\,}{\,}{\,}\left
[b_n,b^{\dag}_m\right ]=\delta_{nm}
\end{equation}
But in the case that (\ref{commutation}) holds true, the commutation relations (\ref{commutation}) are not consistent with the quantization condition (\ref{cancom}). In that case, if we want the quantization condition to hold true, which means that the concept of a particle is consistently defined, then, the commutation relations (\ref{commutation}) no longer hold true, and have to be replaced by \cite{ss,wolfram}
\begin{equation}\label{commutation3}
 [a,a^{\dag} ]=0,{\,}{\,}{\,}{\,}{\,}{\,}{\,}
[b,b^{\dag}  ]=0,{\,}{\,}{\,}{\,}{\,} [ a,a  ]=0,{\,}{\,}{\,}{\,}{\,}{\,}{\,}
 [b,b ] =0, {\,}{\,}{\,}{\,}{\,}{\,}{\,}
[a,b^{\dag} ]=i
\end{equation}
Hence the usual Fock space is equipped with an indefinite metric. However, there is another approach \cite{ss,wolfram}, in which the particle concept of the theory collapses, while the commutation relations (\ref{commutation12}) and (\ref{commutation}) both hold true. This approach, the one adopted by \cite{lima,limasos,lima1}, is what we shall adopt in this paper. According to this approach, the total Fock space $\mathcal{H}$ of quantum states is enriched in reference to the previous case, and has the following product form:
\begin{equation}\label{focjk}
\mathcal{H}= \mathcal{H}_{\mathrm{Fock}}\otimes L^2(x_i,y_i)
\end{equation} 
with $\mathcal{H}_{\mathrm{Fock}}$, the Fock space corresponding to the operators that are related to real eigenvalues, while the $L^2(x_i,y_i)$ part corresponds to operators that are related to imaginary eigenvalues. We shall not go into details, since we are interested in the qualitative features of this theory (for a detailed presentation see \cite{ss,wolfram}). In this theory, the total Hamiltonian of the system, that acts on the states of the total Fock space, is modified to take into account the states with imaginary eigenvalues. The final Hamiltonian contains inverted oscillator-like terms, that contain  the operators $a,b$, and is unbounded from below. Actually, these oscillator-like terms make the Hamiltonian unbounded from below. Furthermore, the quantization conditions (\ref{cancom}), no longer hold true and the particle interpretation of the theory collapses. Recall the similarity with the gravitational case we described previously, in which case there was no particle description and in addition an unbounded from below Hamiltonian, with inverted oscillator-like terms. Let us compute the vacuum energy for the scalar system in external electric field case. The energy momentum tensor corresponding to the
charged scalar is equal to \cite{embacher}:
\begin{equation}\label{energymomentum}
\mathcal{T}_M^{\mu \nu}=\varphi^*\left
(\overleftarrow{\mathcal{D}}^{\mu}\overrightarrow{\mathcal{D}}^{\nu}+\overleftarrow{\mathcal{D}}^{\nu}\overrightarrow{\mathcal{D}}^{\mu}\right
)\varphi-\eta^{\mu \nu}\mathcal{L}_M
\end{equation}
with $\mathcal{D}=\partial^{\mu}-ie\mathcal{A}^{\mu}$. It is easy to find
the vacuum energy for each level ``$n$" (with
$\mathcal{A}^{\mu}(\vec{x})=(-V_0,0,0,0)$),
\begin{equation}\label{vacuumenergylev}
\langle 0_f|\mathcal{T}_{M_{(n)}}^{0
0}|0_f\rangle=\left(\phi_n^2(\vec{x})\lambda_n^2+V_0^2e^2\phi_n^2(\vec{x})+\left(\partial_i\phi_n(\vec{x})\right)^2\right)e^{2\lambda_nt}
\end{equation}
As we can see, each vacuum energy level is exponentially
amplified. The similarity in the two cases (strong electric field
and strong gravitational field) is obvious. Let us briefly recapitulate here what we found. The Hamiltonian of the system contains some inverted oscillator terms \cite{ss}, which correspond to the imaginary eigenfrequencies. For the usual quantum-mechanical state space, this leads to the no-vacuum quantization. The other quantization corresponding to commutation relations (\ref{commutation12}), can only be obtained by abandoning the usual structure of quantum theory, materialized by commutation relations (\ref{commutation12}) and (\ref{commutation}), in favor of an indefinite metric, materialized by the commutation relations (\ref{commutation3}). In our approach (the one adopted by \cite{lima,limasos,lima1} for the gravitational case) the particle concept of the theory is lost. In early works this phenomenon was treated as an instability \cite{ssw,embacher} and
the indefinite metric (\ref{commutation3}) was introduced \cite{ss} in order the problem
leads to physically acceptable particle like solutions. However in
our case, we are not interested in finding bound state solutions.
We studied the implications that the ``adiabatic vacuum"
approximation has on the charged scalar field vacuum expectation
values, when a strong electric field is turned on (assuming flat
spacetime).

\noindent A very interesting question to ask is, what is the dependence of the eigenvalues $\lambda_i$ in terms of the charge, the mass and the depth of the potential $V_0$. This question was addressed numerically in reference \cite{fulling}. In order to have an idea of this dependence, consider an equivalent problem with spherical symmetry \cite{ssw}, for which the electric potential is: 
\begin{equation}
\mathcal{A}_{0}(r)= \Bigg{\{}\begin{array}{c}
   -V_0, {\,}{\,} {\,}{\,}0<r<r_c \\
   \\
  0,{\,}{\,}{\,}{\,}{\,}{\,}r>r_c\\
\end{array}
\end{equation}
with $r_c$ a characteristic radius at which the potential changes. Considering only the spherical symmetric part $l=0$ of the field (with ``$l$'' the angular momentum quantum number) , solving the spherical version of the Klein-Gordon equation, we end up to the following equation \cite{ssw}:
\begin{equation}\label{newadd}
\xi \cos \xi =-r_c\sqrt{(1-\omega_i^2)}
\end{equation}
with $\xi =r_c\sqrt{(V_0+\omega_i)^2-1}$. The numerical study of equation (\ref{newadd}), can yield imaginary values for the eigenfrequencies $\omega_i$ (see reference \cite{ssw}, for a more detailed analysis). Hence equation (\ref{newadd}), provides some information on how the eigenfrequencies depend upon the aforementioned parameters. Notice that, only the parameter $V_0$ (the depth of the potential) determines if the eigenvalues are real or imaginary. So, for sufficiently negative potential, imaginary eigenvalues occur.

It worths to recapitulate the similarities we found for the gravitational vacuum amplification and for the electric vacuum amplification. For simplicity, we refer to these as gravitational amplification (shortened to "GA'') and electric amplification (shortened to ``EA''). According to our findings, the similarities are the following:

\begin{itemize}

\item In both the EA and GA cases, when the potential is deep enough and negative, imaginary eigenfrequencies occur.

 \item In both the EA and GA cases, the particle content of the quantum theory is lost when imaginary eigenfrequencies occur.

\item In both the EA and GA cases, the quantum Hamiltonian is unbounded from below when imaginary eigenfrequencies occur. The term that causes the unboundedness of the Hamiltonian, contains operators that are related to the imaginary eigenfrequencies.

\item In both the EA and GA cases, the particle content and the unbounded Hamiltonian occur when we assume canonical commutation relations for the operators associated to the imaginary eigenfrequencies, namely $[c_{\sigma},c^{\dag}_{\sigma'} ]=\delta_{\sigma\sigma'}$ and $ [a_n,a^{\dag}_m]=\delta_{nm},{\,}{\,}
[b_n,b^{\dag}_m ]=\delta_{nm} $ for the GA and EA cases respectively.

\item In both the EA and GA cases, as a result of the imaginary eigenfrequencies, the vacuum is exponentially amplified. 

\end{itemize}

\noindent The absence of a particle interpretation in the quantum theory at hand, might not be a disadvantage of the theory, if we put the whole theory in the vacuum amplification context of the gravitational analogue, in which case a formal definition of a particle is absent too. It would be interesting to combine strong electric and
gravitational field along the lines of reference \cite{lima,limasos,lima1}. In
fact an early study of such a setup (but the study was focused on
vacuum polarization in black holes) can be found in
\cite{gibbons}. Note that in early studies of quantum field theory under the influence of strong electric
fields, the main interest was in finding stationary states for the
scalar field \cite{ss}. In these studies a stationary state could
not be formally defined unless the field operators satisfy certain
relations \cite{ss,ssw}, namely those of equation (\ref{commutation3}). In the case that a field satisfies the
relations (\ref{commutation}), phenomena similar to the
Klein paradox \cite{fulling} occur. We could say that the deep potential well actually acts as a laser (as noted by D. Sciama in \cite{fulling}, see note added in proof of \cite{fulling}) that amplifies the vacuum, or using the terminology of Lima et al., the deep potential well awakes the vacuum.

\end{document}